\documentclass[preprint]{aastex}
\input epsf

\newcommand{\etal}{{\it et al.\thinspace}}

\slugcomment{Astrophys. J. {\it Letters} (in press)}

\shorttitle{X-Ray Photoabsorption in O VI}
\shortauthors{Pradhan}

\begin{document}
\let\typeset\relax

\title{X-Ray Photoabsorption in KLL Resonances of O VI And Abundance
Analysis}

\author{Anil K. Pradhan}
\affil{Department of Astronomy, The Ohio State University, Columbus, OH
43210}

\begin{abstract}

 It is shown that photoabsorption via autoionizing resonances may 
be appreciable and used for abundance analysis. Analogous to 
spectral lines, the `resonance oscillator strength' $\bar{f}_r$ 
may be defined and
evaluated in terms of the differential oscillator strength df/d$\epsilon$ that
relates bound and continuum absorption. X-ray photoabsorption in KLL
(1s2s2p) resonances of O~VI is investigated using highly resolved relativistic
photoionization cross sections with fine structure. It is found that 
$\bar{f}_r$ is comparable to that for UV dipole transition in O~VI (2s -
2p) and the X-ray ($1s^2 \ ^1S_0 - 1s2p \ ^1P^o_1)$ transition in
O~VII. The dominant O~VI(KLL)
components lie at $\lambda\lambda$ 22.05 and 21.87 $\AA$. These
predicted absorption features should be detectable by
the {\it Chandra X-Ray Observatory} (CXO) and the {\it X-Ray Multi-Mirror
Mission} (XMM). The combined UV/X-ray spectra of
O~VI/O~VII should yield valuable information on the ionization
structure and abundances in sources such as the `warm absorber' region
of active galactic nuclei and the hot intergalactic medium.
Some general implications of resonant photoabsorption are addressed.

\end{abstract}




\section{INTRODUCTION}

 Whereas line absorption has been well studied and used for diagnostics
and abundance analysis (Spitzer 1978), resonant absorption does not appear
to have been similarly considered. This is probably due to
the general complexity of resonances that require
rather elaborate atomic physics calculations. On the other hand,
resonances are ubiquitous, and may considerably  affect the effective
cross sections. In this {\it Letter} the theoretical treatment of
resonant absorption is generalized using the quantity the differential
oscillator strength that describes both bound-bound and bound-free
absorption on either side of the ionization threshold. 
 The method is applied to K-shell X-ray absorption in Lithium-like Oxygen.

 The O~VI UV absorption in the $2s ^2S_{1/2}- 2p ^2P^o_{3,2,1/2}$ transition
at $\lambda\lambda$ 1031.91 and 1037.61  $\AA$ is widely observed 
in sources such as quasars and AGN (Mathur \etal 1994, Tripp \etal 2000),
and {\it Far Ultraviolet Spectroscopic Observer} (FUSE) sources 
(e.g. Savage \etal 2000). Hellsten \etal (1998) have
predicted an `X-ray forest' of O~VII and O~VIII absorption lines 
from the low-z hot intergalactic
medium as a probe of baryonic matter. Recently, X-ray absorption
and emission line spectra have been reported from the CXO from H- and 
He-like ions such as O~VIII and O~VII (Kaspi \etal 2000, Kaastra \etal 2000). 
 In their work Mathur \etal (1994)
reported on UV/X-ray absorption from the same element but in
different ionization states, O~VI and O~VII, from the `warm absorber' 
region of AGN. The possibility
of the same ionic species (O~VI) as both the UV and X-ray absorber is
therefore of further interest for ionization structure and
abundance studies.
It is shown in this
{\it Letter} that resonant K-shell X-ray absorption by O~VI should lie
among, but distinct from,
the prominent emission lines of O~VII due to $2(^3S_1, ^3P^o_{1,2},\
^1P^o_{1}) \longrightarrow 1(^1S_0)$ transitions.

 KLL resonances are normally see as {\it emission} lines (Gabriel 1972). 
Di-electronic
recombination of highly ionized ions, for example $e~+~Fe~XXV \longrightarrow
Fe~XXIV$, leads to di-electronic satellite (DES) lines that are useful
diagnostics of high-temperature sources such as tokamaks and solar
flares (e.g. Bely-Dubau \etal
1982, Beirsdorfer \etal 1992). The radiative decay rates of many DES of Fe~XXV approach or exceed 
autoionization rates (e.g. Pradhan and Zhang 1997). For lighter elements, 
such as Oxygen, radiative decays are much smaller and these resonances should
manifest themselves primarily in absorption, as demonstrated  in this
work.

\section{THEORY AND COMPUTATIONS}

 The differential oscillator strength may be used to relate bound-bound 
and bound-free
absorption as follows (e.g. Seaton 1983, Fano and Rau 
1986, Pradhan and Saraph 1977):

\begin{equation}
\frac{df}{d\epsilon} = \left[ \begin{array}{l}
                         \frac{\nu^3}{2z^2} f_{line} \ \ \ \ , 
\hskip 2cm \epsilon < I \\
                         \frac{1}{4\pi^2 \alpha a_0^2} \sigma_{PI} 
 \ , \hskip 2cm \epsilon > I
                         \end{array} 
                         \right.
\end{equation}

 where $f_{line}$ is the line absorption oscillator strength,
$\sigma_{PI}$ the photoionization cross section, I the ionization
potential, z the ion charge,
$\nu$ the effective quantum number
at $\epsilon = -\frac{z^2}{\nu^2}$ in Rydbergs,
and  $\alpha$ and $a_0$ are the fine structure constant
and the Bohr radius respectively.
 The quantity $\frac{df}{d\epsilon}$ describes the strength of photoabsorption
per unit energy, in the discrete
bound-bound region as well as the continuum bound-free region, 
continuously across the ionization threshold.  We may write,

\begin{equation}
\lim_{n \rightarrow \infty} \left( \frac{\nu_n^3}{2z^2} \right) f (J_i -
J_n) = \lim_{\epsilon \rightarrow 0} \left( \frac{1}{4\pi^2 \alpha
a_0^2} \right) \sigma_{PI} (J_i - \epsilon (J)) ,
\end{equation}

 where $J_i,J_n$ represent the symmetries of the initial and final bound
levels, and $J$ represents the continuum symmetry, governed by the
usual dipole selection rules $\Delta J = 0, \pm 1; \pi \rightarrow - \pi$. 
The photoionization  cross
sections contain Rydberg series of autoionizing resonances converging on
the excited levels of the residual (photoionized) ion. The
effective photoabsorption is generally enhanced in the vicinity of
resonances.

 The $\frac{df}{d\epsilon}$  reflects the
same resonance structure as the $\sigma_{PI}$ in the bound-free continuum.
Combining the two forms of $\frac{df}{d\epsilon}$ we therefore define, 
in the vicinity of a resonance, the integrated `resonance absorption
oscillator strength' as:

\begin{equation}
 \bar{f}_{res} (J_i \longrightarrow J_f)  =  \int_{\Delta E_{res}} \left(
\frac{df (J_i \longrightarrow J_f)}{d\epsilon} \right) d\epsilon  ,
\end{equation}

where $J_i,J_f$ represent the initial bound and the final
continuum symmetries.
 Eq. (3) may be evaluated from the detailed $\sigma_{PI}$ for the
symmetries concerned
provided the resonance profile is sufficiently well delineated. In
practice this is often difficult and elaborate methods need to be
employed to obtain accurate positions and profiles (the background and
the peaks) of resonances.
Relativistic effects need to be included to differentiate the fine
structure components. Using the coupled channel formulation based 
on the R-matrix and the relativistic Breit-Pauli R-matrix (BPRM) method 
(Burke \etal 1971, Berrington \etal 1995) a large number of 
photoionization cross
sections have been calculated for all astrophysically abundant elements 
including resonance structures, particularly in the Opacity Project 
and the Iron Project works (Seaton \etal 1994, Hummer \etal 1993).
The BPRM formulation has been extended to
theoretically self-consistent calculations of photoionization/recombination of 
atomic systems (e.g. Nahar \etal 2000a,b, Zhang \etal 1999), 
including a unified treatment
of total non-resonant {\it and} resonant recombination 
(radiative and di-electronic
recombination). 

Photoionization of, and electron recombination to, 
an atom  is described in terms of the {\it same} eigenfunction expansion
over coupled levels of the residual (`core' or `target') 
ion.
 Recently, BPRM photoionization/recombination calculations have been carried
out for Li-/He-/H- like carbon and iron: C~IV/C~V/C~VI (Nahar \etal 2000a) and
Fe~XXIV/Fe~XXV/Fe~XXVI (Nahar \etal 2000b) for applications to X-ray
photoionization and NLTE modeling.
We similarly consider the photoionization of the ground state of O~VI, 
$1s^22s \ (^2S_{1/2})$ into
all n=1,2,3 fine structure levels of O~VII, $1s^2 (^1S_0), 1s2s
(^3S_1,^1S_0)$, $1s2p (^3P^o_{0,1,2},^1P^o_{1})$, $1s3s (^3S_1,^1S_0), 1s3p
(^3P^o_{0,1,2})$, $1s3d (^3D_{1,2,3}, ^1D_2)$ in the target expansion. 
Thus K-shell photoionization of O~VI $(1s^22s)$, i.e. excitation-autoionization 
via the 1s $\rightarrow$ 2p transition resulting in 1s2s2p (KLL)
resonances, is considered.
We consider the initial bound state of O~VI
($^2S_{1/2}$) with symmetry J = 0.5 (even parity), and final continua
of O~VII with J = 0.5 and 1.5 (odd parity). The KLL resonances of 
interest here are: $1s2p (^3P^o) 2s \ \  [^4P^o_{1/2,3/2}, ^2P^o_{1/2,3/2}]$
and $1s2p (^1P^o) 2s \ \ [^2P^o_{1/2,3/2}]$. The autoionization and
radiative decay rates, and cross sections with and without radiative
decay of resonances, are calculated by analysing the poles in the
complex dipole matrix elements using the method described in (Pradhan and
Zhang 1997). The cross sections are resolved on a very fine mesh of
up to $10^{-6}$ eV.

\section{RESULTS}

 Fig. 1a shows the photoionization cross section of O~VI from the L-shell
(2s) ionization threshold at O~VII $(1s^2 \ ^1S_0)$, up to the K-shell 
ionization thresholds at 1s2s, 1s2p levels of O~VII. Converging on to
the K-shell edges are the KLn $n \geq 2$ complexes of resonances. We
resolve the lowest resonance ``doublet" feature at $E \simeq 41.5$
Rydbergs into 4 fine structure components of
the 1s2s2p complex in detail in Figs. 1b,c. The peak values in Figs.
1b,c are up to 4 order of magnitude higher than the ``edges" in Fig. 1a,
and indicative of the corresponding photoabsorption resonance (PAR) strengths.
We label these as PAR resonances in {\it absorption} to distinguish them 
from the same resonances seen as DES in emission for other Li-like ions
such as Fe~XXIV mentioned above. The computed wavelengths of the two
features in Fig. 1a are $\lambda\lambda$ 22.05 and 21.87 $\AA$, each with
twin J=0.5,1.5 components shown in Figs. 1b,c. (The energy scale is also
given in KeV on top in Fig. 1a).

 It is clearly important to resolve the resonances completely in order
to evaluate the PAR strength $\bar{f}_r$ according to  Eq. (3).
The probability of resonances decaying
radiatively back to the bound state(s) of O~VI, vs. the autoionization
probability, is included using the radiation damping
procedure described in Pradhan and Zhang (1997). 
Although not obvious on the Log scale, the
radiatively damped cross sections (solid lines) are up to 40\% lower
than undamped ones (dashed lines) at peak values. The
resonances in Fig. 1c have no significant damping (dashed and solid
lines merge). The computed resonance positions $E_r$, the 
autoionization and radiative decay rates
$\Gamma_a$ and $\Gamma_r)$, and
the PAR strengths $\bar{f}$ using radiatively damped and undamped cross
sections (the latter in parenthesis) are given in Table 1. 

 Fig. 2 shows the computed differential oscillator strength
$\frac{df}{d\epsilon}$ for O~VI
photoabsorption over a wide energy range, from the 2s-2p transition in
UV, to the X-ray absorption in KLL. The fine structure J = 0.5,1.5 has
been summed over in oscillators strengths and photoionization cross
sections. The BPRM line oscillator strengths for
the discrete ($2s \ ^2S_{1/2} - np \ ^2P^o_{1/2,3/2}$) transitions were
also computed. In accordance
with Eq. (2), there is smooth continuation of $\frac{df}{d\epsilon}$
across the 2s ionization threshold. Eq. (2) provides a
stringent check on both the line oscillator strengths and 
photoionization cross sections for each symmetry.
 
  The relative line and resonance strengths in O~VI are qualitatively
apparent from Fig. 2. Quantitatively, the computed PAR strengths are
given Table 1. Identification of the PAR `satellites' is in accordance
with the standard DES notation (Gabriel 1972), where the KLL resonances
are labeled by letters a-v. The four dominant components of the 1s2s2p complex
according to the calculated $\Gamma_a,\Gamma_r$ are the ones labeled
`t',`s',`r' and `u'. Two weaker components `q' and `v' are not resolved 
since their autoionization rates are about 2 order of
magnitude smaller, and therefore their contribution to photoabsorption
should be negligible (see, for example, the corresponding $\Gamma_a$ for
Fe~XXV in Bely-Dubau \etal 1982, and Pradhan and Zhang 1997). The
calculated $\Gamma_r$ for resonances at $\lambda$ 21.87 $\AA$ are
roughly two orders of magnitude smaller than the $\Gamma_a$, whereas for
the resonances at $\lambda$ 22.05 $\AA$  
$\Gamma_a$ and $\Gamma_r$ are comparable.
That accounts for the significant radiation damping in the latter case
(Fig. 1b).
The computed $\bar{f}_r$ (Eq. 3) for the PAR satellites are found to
be comparable to typical line oscillator strengths $f_{\ell}$ for dipole
transitions. The combined $\bar{f}_r (\lambda 22.05)$ = 0.408 (0.5761), 
and $\bar{f}_r
(\lambda 21.87)$ = 0.0606 (0.0608). By comparison the O~VI f$_\ell$(2s-2p)
for the UV fine structure doublet $\lambda\lambda$ 1031.91 and 1037.61
$\AA$ are 0.199 and 0.066 respectively, and the
O~VII f($1^1S_0 - 2^1P^o_{1}$) is 0.6944 (Wiese \etal 1996); the latter
is nearly equal to the sum of the (undamped) $\bar{f}_r$, as expected. 

\section{DISCUSSION}

 Significant X-ray absorption by O~VI at $\lambda$ 22.05 $\AA$, and a weaker
one at $\lambda$ 21.87 $\AA$, is expected based on the theoretically
computed resonance strengths. 
These wavelengths lie in the range spanned by the emission lines of
O~VII due to electron impact excitation and recombination-cascades (e.g.
Pradhan 1982) in
transitions $2(^3S_1, ^3P^o_{2,1}, ^1P^o_1)
 \longrightarrow 1(^1S_0)$ at $\lambda\lambda$ 22.101, 21.804, and
21.602 $\AA$, usually labeled as `f',`i' and `r' for
forbidden, intercombination and resonance transitions.
 Although the O~VI absorption and O~VII emission features
lie close together, they should be distinguishable with the CXO
resolution (e.g. Canizares \etal 2000).

An inspection of the X-ray spectra of the
Seyfert galaxy NGC 5548, reproduced in Fig. 3 from
Kaastra \etal (2000), appears to show absorption dips at
$\lambda\lambda$ 22.05 and 21.87 $\AA$ (dashed lines), 
both lying in between the `i'
and `f' emission lines of O~VII. Further, the $\lambda$ 22.05
dip is much stronger, as inferred by the $\bar{f}_r$ given in
Table 1. Kaastra \etal(2000) 
do not comment on these features; however, the
combined O~VI absorption might be comparable to the {\it net} absorption
in the resonance `r' line of O~VII (albeit reduced by `r' emission).
Since the O~VII `i' and `f' lines at $\lambda\lambda$ 21.804 and 22.101
$\AA$ are forbidden, with Einstein A-values 1.04 $\times 10^3$
and 3.31 $\times 10^5$ sec$^{-1}$ respectively (Wiese \etal 1996), 
they should not exhibit significant absorption, unlike the `r' line with
A-value of 3.309 $\times 10^{12}$ sec$^{-1}$ which does have an absorption
component (Kaastra \etal 2000). It might be noticed from Fig.3 that
fits to all features are slightly shifted in $\lambda$ due to velocity fields.

 In addition to the KLL PAR's described herein,
the closely spaced KLn ($ 2 < n \leq \infty$) absorption may be discernible as a
pseudo-continuum below
the 1s2$\ell$ K-shell ionization edges; for O~VI (Fig. 1a) these
higher energy features might be between $17.6 - 19.4 \AA$ (0.64 - 0.71 KeV).
The KLn are not fully resolved in Fig. 1a. Being much narrower than the
KLL, since $\Gamma_a \sim n^{-3}$, they are also more likely to be
radiatively damped out, i.e. appear in emission via the DR process. 

 The determination of column densities (N) and ionic abundances using observed
equivalent widths $W_{\lambda}$, and the undamped
PAR strengths $\bar{f}_r$ (Table 1), may be made with
the standard curve-of-growth, i.e. Log $(W_{\lambda}/\lambda)$ vs. Log
$(N \bar{f}_r \lambda)$.
Given that the K-shell resonance absorption strengths are
substantial, we should expect $W_{\lambda}$(O~VI) and N(O~VI)
from X-ray observations to be consistent with those obtained from O~VII (Kaspi
\etal 2000, Kaastra \etal 2000).
The non-resonant background has little effect on the
results; the predominant contribution is from energies
close to the peak resonance values. The total $\frac{df}{d\epsilon}$ in
a given energy range quantifies the effective photoabsorption therefrom.

 The uncertainties in the photoionization calculations should be small.
The BPRM calculations for
photoionization/recombination show excellent agreement in
magnitude, shape, and positions of resonances compared with measured
photo-recombination spectra from ion storage rings: C~IV,C~VI and O~VII
(Zhang \etal 1999), Ar~XIV (Zhang and Pradhan 1997), and Fe~XXV (Pradhan and
Zhang 1997). Nonetheless, the {\it ab initio} BPRM calculations are not 
quite of spectroscopic accuracy. For example, the computed ionization
potential of O~VI is 10.1495 Ryd, compared to the observed value of
10.1516 Ryd -- a difference of 0.02\%, which may be
the uncertainty in the computed resonance positions and wavelengths
in Table 1.

\section{CONCLUSION}

 A few conclusions may be drawn from this study.

1. The strength of resonant photoabsorption may be computed and used for
abundance analysis. Highly accurate and detailed atomic photoionization 
cross sections are required to obtain the corresponding PAR strengths.
2. The X-ray spectra from CXO and XMM should display the PAR absorption
features at $22.05 \AA$ and $21.87 \AA$, lying in between the well known
triplet emission features of He-like O~VII (Fig. 3). 
O~VI and O~VII exist in very different plasma
conditions, with peak temperatures for maximum abundance that may 
differ up to an order of magnitude 
depending on photoionization and/or
coronal equilibrium (Kallman 1995, Arnaud and Rothenflug 1985). 
The O~VI absorption in both UV and X-ray provides an
additional tool for ionization and abundance studies.
An examination of X-ray spectra is suggested
for both the O~VII emission and the O~VI absorption features.
3. Radiative decay rates for autoionizing resonances may be obtained
from the integrated $\frac{df}{d\epsilon}$ since the Einstein
A-values are related to f-values (Wiese \etal 1996).
As confirmation of accuracy of the method presented, the computed $\bar{f}_r$
in Table 1 are nearly equal (to two decimal figures) to those obtained from the
$\Gamma_r$. However, the quantity $\frac{df}{d\epsilon}$ is more general
and represents photoabsorption in lines, resonances, and the
non-resonant background at all energies. As such, it may be useful
in complex cases with many overlapping resonances or lines. 
4. The PAR features in absorption should manifest themselves as
di-electronic satellites (DES) along an iso-electronic sequence 
as $\Gamma_r \sim Z^4$. Contrariwise, unlike heavier elements like 
iron where the DES are strong, for lighter elements like oxygen DES
emission is very weak  (possibly undetectible), and
the PAR satellites could be important absorption line diagnostics.
5. Radiative transfer in resonances may be significant and should be
considered in NLTE and photoionization models.
 These and other points will be discussed in subsequent works.

\acknowledgments

 I would like to thank Jordi Miralda-Escud\'{e} for his suggestion
 regarding K-shell absorption that led to this work, and Smita
Mathur for discussions. I am grateful to Sultana Nahar, Hong Lin Zhang, 
and Justin Oelgoetz for collaboration on BPRM photoionization calculations.
 This work was partially supported by the NSF and the NASA
Astrophysical Theory Program.

\newpage
%-----------------------%
%  references           %
%-----------------------%

\def\r{\leftskip10pt \parindent-10pt \parskip0pt}
\def\apj{ApJ}
\def\apjs{ApJS}
\def\apjl{ApJL}
\def\aj{Astron. J}
\def\pasp{Pub. Astron. Soc. Pacific}
\def\mn{MNRAS}
\def\aa{A\&A}
\def\aasup{A\&A Suppl.}
\def\baas{Bull. Amer. Astron. Soc.}
\def\jqsrt{J. Quant. Spectrosc. Radiat. Transfer}
\def\jpb{Journal of Physics B: Atom. Molec. Opt. Phys.}
\def\pra{Physical Review A}
\def\adndt{Atomic Data And Nuclear Data Tables}

%\newpage

\begin{table}
\caption{Calculated parameters for the photoabsorption resonances (PAR) 
in O~VI}
\begin{tabular}{lllcll}
\hline
Identification & $\lambda_{calc}(\AA)$ & E$_r$(KeV) & $\bar{f}_r$ &
$\Gamma_a$ (Ryd,sec$^{-1}$) & $\Gamma_r$ (Ryd,sec$^{-1}$) \\
\hline
$1s2p(^1P^o)2s \ ^2P^o_{1/2}$ (r) & 22.05 & 0.56227 & 0.1410(0.1924) & 3.11(-4),
6.42(+12)& 1.26(-4),2.60(+12)\\
$1s2p(^3P^o)2s \ ^4P^o_{3/2}$ (u) & 22.05 & 0.56231 & 0.2670(0.3837) & 2.76(-4),
5.70(+12)& 1.28(-4),2.65(+12)\\
\hline
\multicolumn{6}{c} {$ < \bar{f}_r  (\lambda 22.05) > $ = 0.408 (0.5761)}  \\     
\hline
$1s2p(^3P^o)2s \ ^2P^o_{1/2}$ (t) & 21.87 & 0.56696 & 0.0216(0.0217) &
3.41(-3),7.11(+13) & 1.46(-5), 3.01(+11)\\
$1s2p(^3P^o)2s \ ^2P^o_{3/2}$ (s) & 21.87 & 0.56700 & 0.0390(0.0391) &
3.43(-3), 7.09(+13) & 1.32(-5), 2.72(+11)\\

\hline
\multicolumn{6}{c} {$< \bar{f}_r (\lambda 21.87) > = 0.0606 (0.0608)$} \\     
\hline
\end{tabular}
\end{table}

\newpage

\figcaption[fig1.eps] {Photoionization cross section of O~VI (a). The
KLL resonance complexes at $\lambda\lambda$ 22.05 and 21.87 $\AA$ are
resolved in (b) and (c) including the fine strcture J-components. 
Note the different energy and cross section
scales: break at 40 Ryd in (a), and Log$_{10} \sigma$ in (b,c).
The resonance peaks in (b,c) are up to 4 orders of magnitude 
higher than in (a). \label{fig1}}

\figcaption[fig2.eps]{The differential oscillator strength 
$\frac{df}{d\epsilon}$ ((summed over 
fine structure) for bound-bound and bound-free photoabsorption in
O~VI from the lowest energy UV transition 2s-2p at $\lambda 1034 \AA$,
to the predicted X-ray PAR transitions at $\lambda\lambda$ 22.05 and
21.87 $\AA$.  \label{fig2}}

\figcaption[fig3.eps]{X-ray emission spectra and absorption spectra of
O~VII from the Seyfert galaxy NGC 5548 (solid lines, Kaastra
\etal 2000). The dashed lines have been added at the predicted
O~VI absorptions features $\lambda\lambda$ 21.87 and 22.05. \label{fig3}}

\end{document}